\documentclass{aa}
\usepackage{graphicx}
\usepackage{txfonts}
\usepackage{natbib}
\begin{document}

\newcommand{\Mr}{\mathcal{M}_{\rm r}}
\def\lapp{\mathbin{\raise1.5pt \hbox{$<$} \hskip-7pt \lower3pt
\hbox{$\sim$}}}
\def\gapp{\mathbin{\raise1.5pt \hbox{$>$} \hskip-7pt \lower3pt
\hbox{$\sim$}}}

\title{On the linear theory of Kelvin-Helmholtz instabilities of relativistic
       magnetohydrodynamic planar flows}

   \author{Z. Osmanov\inst{1}\fnmsep\thanks{something}
          A. Mignone\inst{1,2},
          S. Massaglia\inst{1},
          G. Bodo\inst{2} and
          A. Ferrari\inst{1}
          }

   \offprints{Z. Osmanov}

   \institute{Dipartimento di Fisica Generale, Universit\'a degli Studi di Torino,
              via Pietro Giuria 1, 10125 Torino\\
              \email{name1@to.infn.it}
         \and
INAF/Osservatorio Astronomico di Torino, Strada Osservatorio 20,
           10025 Pino Torinese, Italy
             \email{name2@oato.inaf.it}
             }

   \date{Received ; accepted }


  \abstract
  {}
   {We investigate the linear stability properties of the plane interface separating two
    relativistic magnetized flows in relative motion.
    The two flows are governed by the (special) relativistic equations
    for a magnetized perfect gas in the infinite conductivity approximation.}
   {By adopting the vortex-sheet approximation, the relativistic magnetohydrodynamics
    equations are linearized around the equilibrium
    state and the corresponding dispersion relation is derived and discussed.
    The behavior of the configuration and the regimes of instability are investigated
    following the effects of four physical parameters, namely: the flow velocity,
    the relativistic and Alfv\'enic Mach numbers and the inclination of
    the wave vector on the plane of the interface.}
   {From the numerical solution of the dispersion relation, we find
    in general two separate regions of instability, associated respectively
    with the slow and fast magnetosonic modes.
    Modes parallel to the flow velocity are destabilized only for
    sufficiently low magnetization.
     For the latter case, stabilization is attained, additionally, at
     sufficiently large relativistic velocities between the two flows in relative motion.}
   {The relevance of these results to the study of the stability of astrophysical jets is briefly commented.}

   \keywords{Kelvin-Helmholtz instability -
             relativistic MHD - plasma physics}

  \authorrunning{Osmanov et al.}
  \titlerunning{Kelvin Helmholtz Instability in Relativistic Magnetized Flows}
   \maketitle

\section{Introduction}
%
%
%
%
%
%
The Kelvin-Helmholtz instability (KHI henceforth), acting at the
contact interface between two flows in relative motion, plays a
dynamically important role in a number of different astrophysical
scenarios such as, for example accretion disks, planetary
magnetospheres, stellar and extragalactic jets. It is a classical
instability in fluid dynamics, already discussed at the end of
nineteenth century in the works of Von Helmholtz \& Monats
\cite{hel} and Lord Kelvin \cite{kel}. The classical results of
the linear analysis for incompressible flows separated by a planar
vortex sheet, both in the presence of magnetic fields and for
different geometries are summarized in Chandrasekhar's monograph
\citep{chandra}. They have been later generalized for compressible
case both in the pure hydrodynamical limit \citep[see e.g.][]{ger,
sen64} and in the magnetohydrodynamic (MHD) limit \citep[see
e.g.][]{sen63, PK83} extending the study to the supersonic regime.
In particular it was shown that in the hydrodynamic case no
unstable longitudinal (with respect to the plane interface) modes
are allowed for flow Mach numbers larger than $\sqrt{8}$, while
transverse modes are always unstable.

More recently Choudhury \& Lovelace \cite{roy} studied the linear
instability behavior in the case of a finite-thickness boundary
layer, as a function of the  Alfv{\`e}n velocity and inclination of
the wave vector with respect to the interface. Their main result was
the proof of the existence of two distinct regimes of instability in
the wavevector-Mach number plane corresponding to standing and
traveling MHD waves, the extent of the instability regions depending
upon the magnetization parameter. Magnetohydrodynamic flows in
cylindrical and slab geometries have been studied by Ferrari et al.
\cite{fer81}, \cite{fer82} showing the existence of a new range of
unstable modes, arising for high Mach number flows, due to standing
waves created by reflections at the boundaries of the structures.


For astrophysical problems involving high-energy phenomena relativistic effects become important
either because the relative velocities are close to the speed of light or because the plasma
is relativistically hot or because the magnetic field is very strong
and the Alfv\`en speed becomes close to the speed of light.
For the non magnetic case a relativistic extension of the KHI study was
considered, for a plane parallel velocity discontinuity, by
Turland \& Scheuer \cite{tur}, Blandford \& Pringle \cite{bln} and more recently by Bodo et al.
\cite{bodo04}. In particular Bodo et al. \cite{bodo04} showed that the
dispersion relation can be solved analytically in a frame of reference where the media in the two half-spaces
have equal and opposite velocities and that the
stability criteria have the same form of classical ones, simply substituting the Mach number with the
relativistic Mach number, that was first introduced by K\"onigl \cite{kon80}.

The relativistic MHD case still lacks of a full stability analysis and this is
what we present in this paper. Early attempts were carried out by Ferrari et
al. \cite{fer}. Their analysis, however, assumed the limit of Alfv{\`e}n velocity much smaller than the speed
of light, thus allowing to neglect displacement currents, while the speed of the flow could reach relativistic
values. The main result of this study was the confirmation that KHI disappears when the
Alfv{\`e}n velocity approaches or becomes larger that the speed of sound and when a large density contrast
exists between the two media.

Relativistic magnetized flows can be found in astrophysical
environments associated to AGN jets, to jets from galactic X-ray
binaries, such as GRS 1915+105 (Mirabel \& Rodriguez 1999), and to
pulsar wind nebulae (PWNe). Applications of the Kelvin-Helmholtz
instability mechanism to the case of AGN and galactic microquasar
jets is widely discussed in the literature, mainly for cylindrical
geometries, such as the recent analysis by Hardee \cite{hardee07},
who studied axially magnetized cylindrical relativistic jet
embedded in a magnetized sheath. In addition KHI appear to be the
physical mechanism to slow down relativistic jets by producing
entrainment of the ambient medium. Bodo et al. \cite{bodo03},
Rossi et al. \cite{rossi04} and Rossi et al. (2008) have studied
the nonlinear evolution of KHI in cylindrical relativistic jets to
explain the dichotomy of FRI - FRII extragalactic radiosources.
Instead the planar case can apply to the case of PWNe, when a
relativistically hot magnetized plasma forms where the pulsar
ultra-relativistic wind interacts with the surrounding supernova
ejecta. The linear stability analysis that we present here will be
relevant for understanding the physical phenomena involved in the
destabilization process.

In this work we present, for the first time to our knowledge,
a complete treatment of the linear stability problem in
the plane-parallel geometry, by considering the full system of
equations (i.e. without neglecting the displacement current and
accounting for non relativistic as well as relativistic Alfv{\`e}n
velocities). It is shown that the stability properties depend on
four physical parameters which we conveniently
choose to be the relativistic Mach number, the flow velocity, the
Alfv{\`e}n speed and the inclination of the projection of the wave
vector onto the plane of the interface.

The paper is organized as follows. In \S\ref{sec:disprel} we derive
the general dispersion relation, in \S\ref{sec:results} we present and discuss the
results for the non-relativistic and relativistic cases, and we add some astrophysical comments
 in \S\ref{sec:summary}.

\section{Relevant Equations}
%
%
%
%
%
%

\subsection{The Equations of Relativistic MHD}
%
%
%

The equations of relativistic magnetohydrodynamics (RMHD henceforth)
govern the evolution of a (special) relativistic magnetized fluid.
They can be cast as a system of conservation laws, describing
energy-momentum and mass conservation:
\begin{equation}\label{eq:tensor}
  \nabla_\mu T^{\mu\nu} = 0 \,,\quad
  \nabla_\mu \left(\rho u^\mu\right) = 0 \,,
\end{equation}
where $\rho$ is rest mass density and
$u^\mu\equiv \left(\gamma, \gamma\vec{v}\right)$
is the fluid four-velocity ($\gamma \equiv$ Lorentz factor,
$\vec{v}\equiv$ three velocity) such that $u^\mu u_\mu = -1$.
The speed of light is set to unity everywhere.
The expression of the energy-momentum tensor can be found under the physical
assumptions of constant magnetic permeability and infinite conductivity,
appropriate for a perfectly conducting fluid
(see, for example, the books by Anile \cite{anile} and Lichnerowitz \cite{lichne}).

Under these conditions, the stress energy tensor
for a perfect fluid interacting with an electromagnetic field decomposes
into $T^{\mu\nu} = T^{\mu\nu}_{\rm FL} + T^{\mu\nu}_{\rm EM}$,
where the fluid (FL) and electromagnetic (EM) contributions are given,
respectively, by
\begin{equation}\label{eq:Tem}
 T^{\mu\nu}_{\rm FL} = \rho h\, u^\mu u^\nu + p \eta^{\mu\nu}
 \,,\quad
 T_{\rm EM}^{\mu\nu} = F^\mu_{\; \beta} F^{\nu\beta} -
                       \frac{1}{4}\eta^{\mu\nu} F_{\alpha\beta}F^{\alpha\beta} \,.
\end{equation}
Here $h$ and $p$ are used to denote the gas specific
enthalpy and the thermal pressure, while
$\eta^{\mu\nu} = {\rm diag}\left(-1,1,1,1\right)$ is the
Minkowski metric tensor. For ease of notation, factors like $4\pi$
have been set to unity in the derivation.

The electromagnetic field tensor $F^{\mu\nu}$
appearing in the definitions of the stress energy tensor, Eq. (\ref{eq:Tem}),
obeys Maxwell's equations
\begin{equation}\label{eq:maxwell}
 \partial_{\rm [\alpha}F_{\mu\nu]} = 0 \,,\quad
 \nabla_\mu F^{\mu\nu} = - J^\nu \,,
\end{equation}
where $[...]$ denotes anti-symmetrization and $J^\nu$ is the charge
four-current.
In the limit of infinite conductivity, the rest frame electric field
vanishes identically and the electromagnetic field tensor
becomes orthogonal to the fluid four-velocity, i.e. $F^{\mu\beta} u_\beta = 0$.
Note that this is the only approximation introduced in ideal RMHD and
it becomes identical to the well known non-relativistic expression
$\vec{E} = - \vec{v}\times\vec{B}$, where $\vec{E}$
is the electric vector field.
Unlike classical MHD, however, the displacement current is not discarded in RMHD
and explicitly enters through the definition of the current, given by the second
of Eqs. (\ref{eq:maxwell}).
The high conductivity limit allows to write
$F^{\mu\nu} = \epsilon^{\alpha\beta\mu\nu}b_\mu u_\nu$,
where $b^\mu$ is the magnetic induction four-vector:
\begin{equation}
 b^\mu = \Big[\gamma\vec{v}\cdot\vec{B} , \frac{\vec{B}}{\gamma}
               + \gamma\left(\vec{v}\cdot\vec{B}\right)\vec{v}\Big]\,,
\end{equation}
with $\vec{B}$ and $\epsilon^{\alpha\beta\mu\nu}$ denoting the rest frame magnetic
field and the Levi-Civita pseudo-tensor, respectively.

For the purpose of our derivation, we explicitly rewrite Eq. (\ref{eq:tensor})
in components:
\begin{equation}
\label{en-mom} \frac{\partial\vec{U}}{\partial t} + \sum_{i =
1}^{3}\frac{\partial\vec{f}^i}{\partial x^i} = 0 \,,
\end{equation}
where $\vec{U}$ and $\vec{f^i}$
are the vector of conserved variables and corresponding fluxes:
\begin{equation}\label{u}
  \vec{U} = [\rho\gamma, ~w_t\gamma^2v^j - b^0b^j,
                                 ~w_t\gamma^2    - b^0b^0 - p_t]^T \,,
\end{equation}
\begin{equation}\label{f}
\vec{f^i} = [\rho\gamma v^i,
                       ~w_t\gamma^2 v^iv^j - b^ib^j + p_t\delta^{ij},
                       ~w_t\gamma^2 v^i    - b^0b^i]^T  \,,
\end{equation}
where $p_t$ and $w_t$ are, respectively the total pressure and
enthalpy, expressed as the sum of thermal and magnetic contributions:
\begin{equation}\label{tpres}
p_t = p + \frac{1}{2}|b|^2 \,,\quad
w_t = \rho h + |b|^2  \,,
\end{equation}
where $|b|^2 = b^\alpha b_\alpha$.
Magnetic field evolution is given by the induction equation in Maxwell's
law:
\begin{equation}
\label{ind} \frac{\partial\vec{B}}{\partial t} -
\vec{\nabla}\times (\vec{v}\times\vec{B})  = 0 \,,
\end{equation}
which conserves the same form as in the classical case.
Proper closure is provided by specifying an equation of state
which we take as the constant $\Gamma$-law:
\begin{equation}\label{enth}
 h = 1 + \frac{\Gamma}{\Gamma-1}\frac{p}{\rho}\,,
\end{equation}
where $\Gamma$ is the polytropic index of the gas.

\subsection{Dispersion Relation} \label{sec:disprel}
%
%
%

Our setup consists of a planar vortex sheet interface in the $xz$ plane
at $y=0$ separating two uniform flows moving in the $x$ direction with
opposite velocities $\vec{v}(y>0) = \beta\vec{i}$ and
$\vec{v}(y<0) = -\beta\vec{i}$. The fluids have equal density
and pressure and are threaded by a uniform longitudinal magnetic
field along the direction of relative motion, $\vec{B} = B_0\vec{i}$.

We start our analysis by introducing small deviations around the
equilibrium state. This is better achieved by working in the
rest-frames of the fluids, where perturbations are
sought in the form:
\begin{equation}
\tilde{\vec{B}} =
\tilde{\vec{B}}_0+\tilde{\vec{B}}'+... \,,
\end{equation}
\begin{equation}
\tilde{\vec{v}} = \tilde{\vec{v}}'+... \,,
\end{equation}
\begin{equation}
\tilde{\rho} = \tilde{\rho}_0+\tilde{\rho}'+... \,,
\end{equation}
\begin{equation}
\tilde{p} = \tilde{p}_0+\tilde{p}'+... \,,
\end{equation}
where the tilde denotes quantities in the rest frame. The perturbed
terms of the first order can be expressed by the following:
\begin{equation}
\tilde{\Psi}_{\pm}' \propto
\exp\left[i\left(\tilde{k}_{\pm}\tilde{x} +
\tilde{l}_{\pm}\tilde{y}+\tilde{m}_{\pm}\tilde{z}-
\tilde{\omega}_{\pm}\tilde{t}\right)\right]
\,,
\end{equation}
where $\tilde{\Psi}_{\pm}' =
(\tilde{\vec{B}}'_{\pm},\tilde{\vec{v}}'_{\pm},\tilde{\rho}'_{\pm},
\tilde{p}'_{\pm})$. Subscripts $+$ and $-$ correspond to the
regions $y>0$ and $y<0$, respectively. Wave numbers and frequency
are denoted with $\tilde{k}_{\pm}$, $\tilde{l}_{\pm}$, $\tilde{m}_{\pm}$
and $\tilde{\omega}_{\pm}$.

Proper linearization of the equations leads to the dispersion
relation (Komissarov \cite{kom}):

$$\left(\tilde{\omega}_{\pm}^2-\tilde{k}_{\pm}^2V_A^2\right)\times$$
\begin{equation}
\label{firstdisp}\times\left(\frac{\tilde{\omega}_{\pm}^4}{(\tilde{k}_{\pm}^2+\tilde{l}_{\pm}^2+\tilde{m}_{\pm}^2)^2}+
\frac{\tilde{\omega}_{\pm}^2\tilde{\mu}_{\pm}}{\tilde{k}_{\pm}^2+\tilde{l}_{\pm}^2+\tilde{m}_{\pm}^2}+\tilde{\nu}_{\pm}\right)=0,
\end{equation}
\begin{equation}
 \tilde{\mu}_{\pm} =
C_{\mathrm{s}}^2+V_{\mathrm{A}}^2-C_{\mathrm{s}}^2V_{\mathrm{A}}^2\frac{\tilde{k}_{\pm}^2+
\tilde{l}_{\pm}^2}{\tilde{k}_{\pm}^2+
\tilde{l}_{\pm}^2+\tilde{m}_{\pm}^2},
\end{equation}
\begin{equation}
\tilde{\nu}_{\pm} =
C_{\mathrm{s}}^2V_{\mathrm{A}}^2\frac{\tilde{k}_{\pm}^2}{\tilde{k}_{\pm}^2+
\tilde{l}_{\pm}^2+\tilde{m}_{\pm}^2},
\end{equation}
\begin{equation}
\label{soundsp}
 C_{\mathrm{s}} =\sqrt{\frac{\Gamma p_0}{\rho_0 h_0}},
\end{equation}
\begin{equation}
\label{alvendsp}
 V_{\mathrm{A}} =\frac{B_0}{\sqrt{\rho_0 h_0+B_0^2}},
\end{equation}
which allow the propagation of the  Alfv{\`e}n mode
\begin{equation}
\label{alfv} \frac{\tilde{\omega}_{\mathrm{A}\pm}^2}{\tilde{
k}_{\mathrm{A}\pm}^2} = V_{\mathrm{A}}^2\,,
\end{equation}
and the two magneto-acoustic modes,  (i) slow magnetosonic
\begin{equation}
\label{ssonic}
\frac{\tilde{\omega}_{\mathrm{s}\pm}^2}{\tilde{k}_{\mathrm{s}\pm}^2+\tilde{l}_{\mathrm{s}\pm}^2+\tilde{m}_{\mathrm{s}\pm}^2}
=  \frac{1}{2}\tilde{\mu}_{\mathrm{s}\pm}-
\frac{1}{2}\left[\tilde{\mu}_{\mathrm{s}\pm}^2
-4\tilde{\nu}_{\mathrm{s}\pm}^2\right]^{1/2},
\end{equation}
and (ii) fast magnetosonic
\begin{equation}
\label{fsonic}
\frac{\tilde{\omega}_{\mathrm{f}\pm}^2}{\tilde{k}_{\mathrm{f}\pm}^2+\tilde{l}_{\mathrm{f}\pm}^2+\tilde{m}_{\mathrm{f}\pm}^2}
=  \frac{1}{2}\tilde{\mu}_{\mathrm{f}\pm}+
\frac{1}{2}\left[\tilde{\mu}_{\mathrm{f}\pm}^2
-4\tilde{\nu}_{\mathrm{f}\pm}^2\right]^{1/2}.
\end{equation}
Here $C_{\mathrm{s}}$ and $V_{\mathrm{A}}$ are the sound
speed and the Alfv{\`e}n speed respectively. Notice that both
$C_{\mathrm{s}}$ and $V_{\mathrm{A}}$ are invariant under Lorentz
boosts in the $x$-direction.


In order to obtain the dispersion relation in the laboratory
frame, one has to transform all the quantities by mean of a
Lorentz transformation:
\begin{equation}
\label{om} \tilde{\omega}_{\pm} = \gamma(\omega\mp k\beta)\,,
\end{equation}
\begin{equation}
\label{k} \tilde{k}_{\pm} = \gamma(k\mp \omega\beta)\,,\quad
\tilde{l}_{\pm} =l_{\pm}\,,\quad
\tilde{m}_{\pm} =m \,,
\end{equation}
so that a generic perturbation in our original frame can be
expressed as
\begin{equation}
\Psi_{\pm}' \propto \exp\left[i\left(kx + l_{\pm}y+mz-\omega
t\right)\right]\,.
\end{equation}

By substituting Eqs. (\ref{om})) and (\ref{k}) into Eq.
(\ref{alfv}) we derive immediately that the possible roots are always
real, hence we may already conclude that the Alfv{\`e}n mode will not contribute
to the instability. In the case of Eqs. (\ref{ssonic}) and (\ref{fsonic}) we see instead
that the resulting slow and fast magnetosonic modes
include both real and imaginary roots, and therefore we expect
instability in the parameter domains corresponding to these modes.

Direct substitution into the induction equation (\ref{ind}) and
into the $y$ and $z$ components of the momentum equation in
(\ref{en-mom}) yields
\begin{equation}\label{1y}
B_0\frac{\tilde{k}_{\mp}}{\gamma}B'_{y\pm}+\left[\omega
B_0^2+\gamma\rho_0 h_0
\tilde{\omega}_{\mp}\right]v'_{y\pm}-l_{\pm}(p'_{\pm}+B_0B'_{x\pm})
= 0\,,
\end{equation}
\begin{equation}\label{1z}
B_0\frac{\tilde{k}_{\mp}}{\gamma}B'_{z\pm}+\left[\omega
B_0^2+\gamma\rho_0 h_0
\tilde{\omega}_{\mp}\right]v'_{z\pm}-m(p'_{\pm}+B_0B'_{x\pm})
= 0\,,
\end{equation}
\begin{equation}
(\omega\mp k\beta)B'_{x\pm}-B_0(l_{\pm}v'_{y\pm}+mv'_{z\pm}) =
0\,,
\end{equation}
\begin{equation}
(\omega\mp k\beta)B'_{y\pm}+kB_0v'_{y\pm} = 0\,,
\end{equation}
\begin{equation}
(\omega\mp k\beta)B'_{z\pm}+kB_0v'_{z\pm} = 0\,,
\end{equation}
where $\tilde{k}_{\mp}$ and $\tilde{\omega}_{\mp}$ are
given by Eqs. (\ref{om}, \ref{k}).

With the aid of Eq. (\ref{tpres}) one may express the total pressure
perturbation as:
\begin{equation}
p^t_{\pm} = p_{\pm}+\frac{l_{\pm}B_0^2v'_{y\pm}}{\omega\mp k\beta
},
\end{equation}
and imposing the displacement matching condition at the interface:
\begin{equation}
\label{inmatch} \frac{v'_{y+}}{\omega- k\beta} =
\frac{v'_{y-}}{\omega+ k\beta}\,.
\end{equation}
and the total pressure continuity ($p^t_{+} = p^t_{-}$) at
the interface, one obtains (together with Eq. \ref{1y})
\begin{equation}
\label{first} \frac{l_{+}}{l_{-}} = \frac{\rho_0 h_0
\gamma^2(\omega-k\beta)^2+(\omega^2-k^2)B_0^2} {\rho_0 h_0
\gamma^2(\omega+k\beta)^2+(\omega^2-k^2)B_0^2}\,.
\end{equation}
Introducing the dispersion relation of slow and fast
magnetosonic waves (see the right bracket of Eq.
(\ref{firstdisp})) and solving for $l^2_\pm$ one
gets
\begin{equation}\label{second}
l_{\pm}^2  = -
m^2+\frac{\tilde{\omega}_{\mp}^2\left[(C_{\mathrm{s}}^2+V_{\mathrm{A}}^2)\tilde{k}_{\mp}^2
-\tilde{\omega}_{\mp}^2\right]-C_{\mathrm{s}}^2V_{\mathrm{A}}^2\tilde{k}_{\mp}^4}
{C_{\mathrm{s}}^2V_{\mathrm{A}}^2
(\tilde{k}_{\mp}^2+\tilde{\omega}_{\mp}^2)-(C_{\mathrm{s}}^2+V_{\mathrm{A}}^2)\tilde{\omega}_{\mp}^2}
 \,,
\end{equation}
which, combined with Eq. (\ref{first}), gives the desired dispersion
relation of modes connecting through the planar interface.

A number of considerations can be immediately drawn.
\begin{itemize}
\item
The dispersion relation obtained by substitution of Eq. (\ref{first})
into Eq. (\ref{second}) is
an 8-th degree polynomial in $\omega$. However, only 4
out of a total of 8 complex roots do actually satisfy Eq. (\ref{first}) and
thus are physically significant.
\item
Roots with positive imaginary part of $\omega$ identify unstable modes.
The growth rate will be more conveniently expressed by the imaginary
part of the dimensionless quantity
\begin{equation}
 \phi\equiv
 \frac{\omega}{C_{\mathrm{s}} \sqrt{k^2+m^2}} \,;
\end{equation}
\item
In the non-relativistic limit (i.e. $\beta\to 0$, $\rho h \gg p,
B_0^2$ and $h \to 1$) our dispersion relation reduces to the form
given by Choudhury \& Lovelace \cite{roy} in their equations
(7) and (A2a);
\item
In the limit of vanishing magnetic field
Eqs. (\ref{first}) and (\ref{second}) simplify considerably and the
dispersion relation reduces to the one derived by Bodo et al.
\cite{bodo04} (see their Eqs. (\ref{inmatch}) and (\ref{second}));

\item
For the general case the system of equations
Eqs. (\ref{first}, \ref{second}) has to be solved numerically and these results are discussed in the following sections.

\end{itemize}

\section{Results} \label{sec:results}
%
%
%
%
%
%

Our study will examine the instability dependence on 4 parameters: 1) the fluid velocity
$\beta$ (alternatively, the Lorentz factor $\gamma$), 2) the relativistic Mach number
\begin{equation}
\label{Mach}
 \Mr = \frac{\beta}{C_{\mathrm{s}} } \frac{\sqrt{1-C_{\mathrm{s}}^2}}{\sqrt{1 - \vec{\beta}^2}}\;,
\end{equation}
3) the Alfv{\'e}nic Mach number $\zeta
= V_{\mathrm{A}}/C_{\mathrm{s}}$ and 4) the ratio $f= m/k$, that gives
the angle between the wave number projection on the $xz$ plane and
the flow velocity. In general, relativistic effects come
into play whenever one of $\beta$, $\Mr$ or $\zeta$ (or
a combination of them) describes situations of high fluid
velocities, hot gas or strong magnetic field, respectively.
To this purpose, it is useful to use the explicit relation linking
the Alfv{\`e}n speed and the other parameters:

\begin{equation}
\label{al_vel} V_{\mathrm{A}}=\frac{\beta
\zeta}{\sqrt{\beta^2+\Mr^2 (1-\beta^2)}} \;.
\end{equation}

In the following, we will discuss the non-relativistic case
and subsequently the relativistic case, both
for oblique propagation ($f \neq 0$) and
for propagation parallel to the velocity interface ($f=0$).
The latter case will be useful for the comparison with the results
obtained by previous investigators.

\subsection{Non-relativistic flows} \label{sec:nonrel}
%
%
%
%
%
%

\begin{figure}
\resizebox{\hsize}{!}{\includegraphics{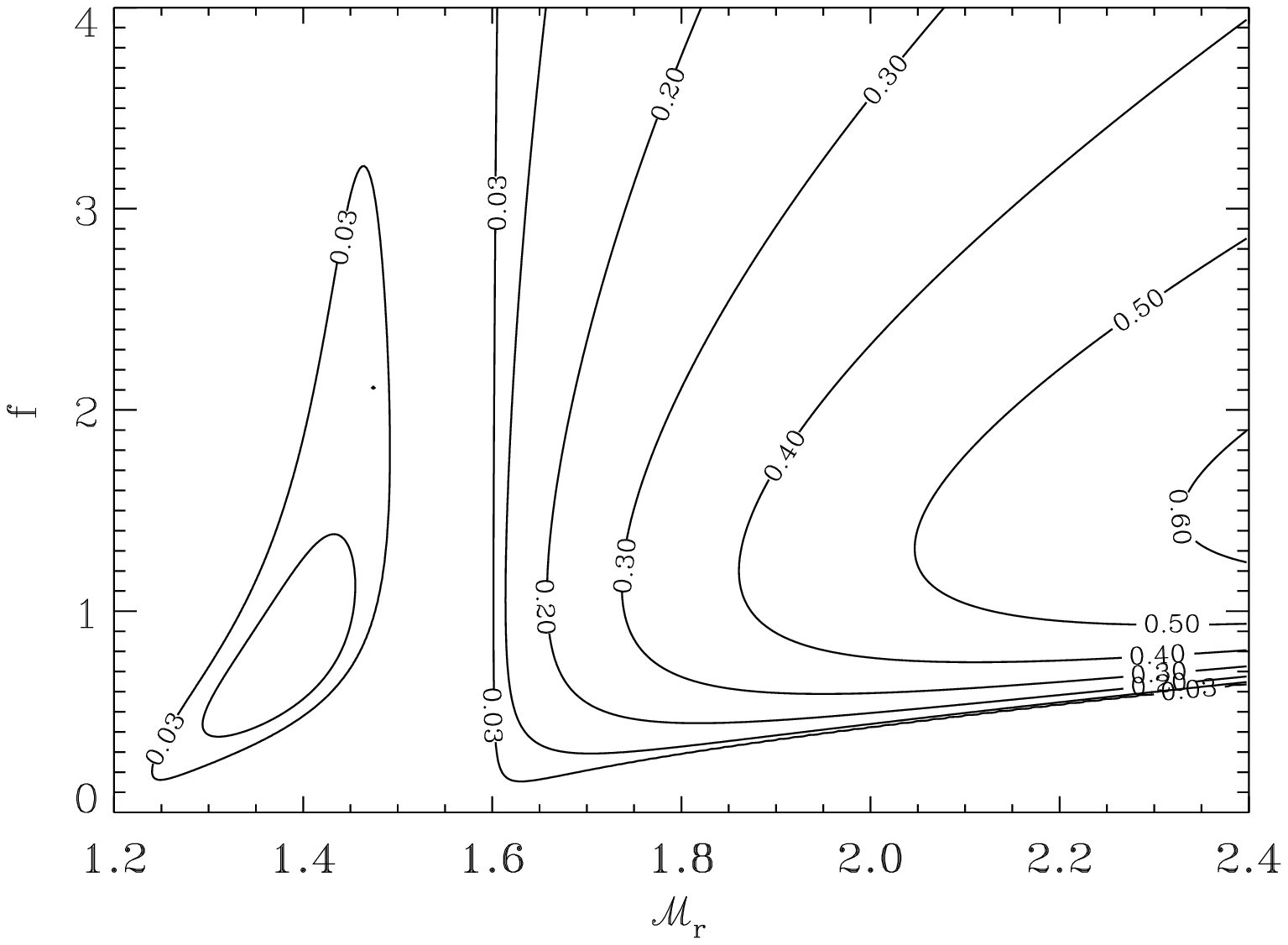}}
\resizebox{\hsize}{!}{\includegraphics{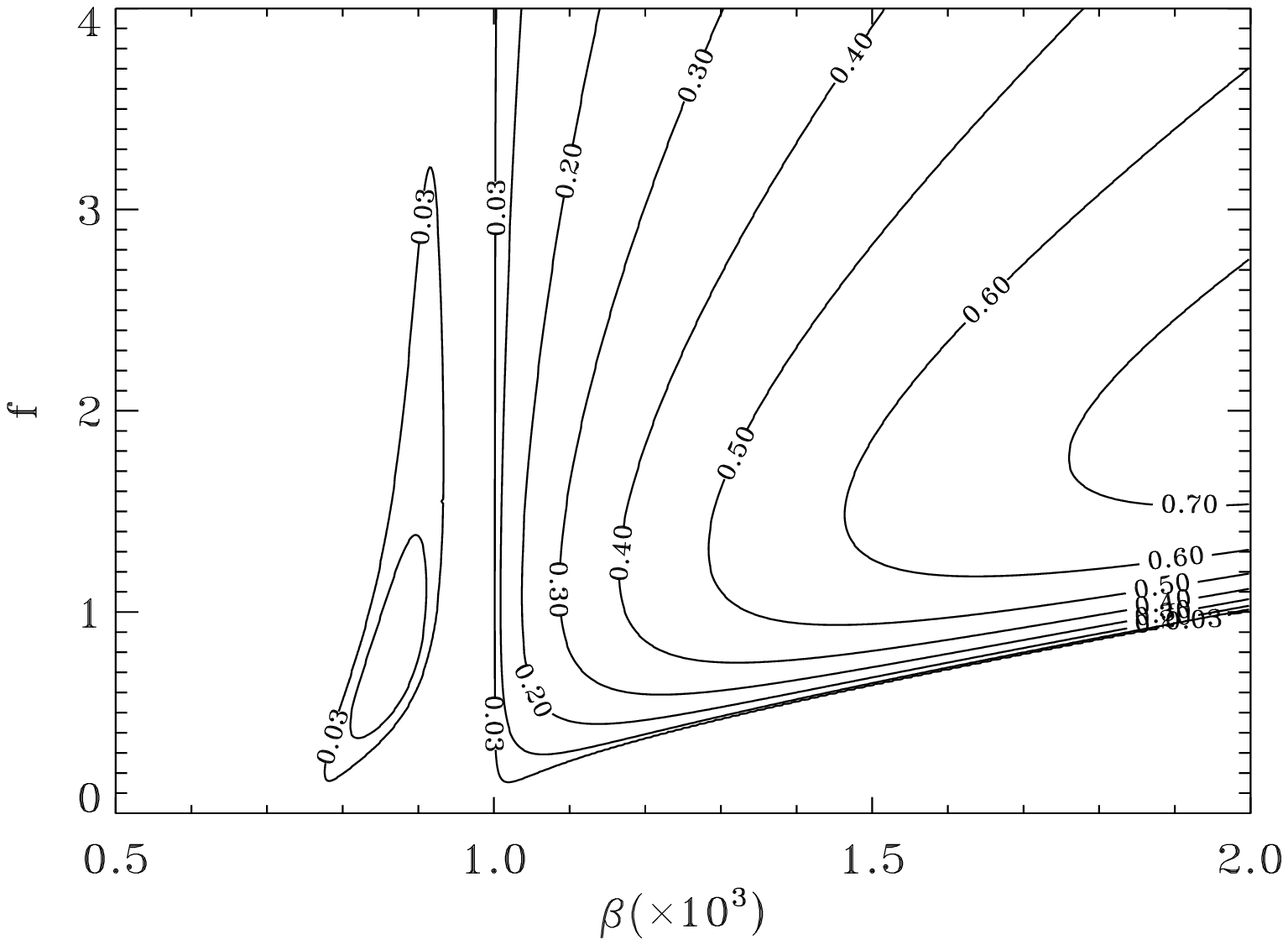}}
 \caption{Two contour plots are shown. Top panel: the contour plot of constant $\mathrm{Im}(\phi)$
surfaces as a function
          of the inclination factor $f$ and relativistic Mach number
          $\Mr$ for fixed Alfv{\'e}nic Mach number $\zeta =1.6$
          and flow velocity $\beta = 10^{-3}$. Bottom panel: the contour plot of constant
$\mathrm{Im}(\phi)$ surfaces as a function
          of the inclination factor $f$ and $\beta$ for the fixed Alfv{\'e}n speed
($V_{\mathrm{A}} = 10^{-3}$) and the sound speed
($C_{\mathrm{s}} = 6.3 \times 10^{-4}$). Here $\Gamma = 5/3$ is used.}
 \label{cont001}
\end{figure}

As a starting point, let us now consider the case of a non-relativistic
fluid, with $\beta = 10^{-3}$ and $\Gamma = 5/3$.

A convenient way to gain a comprehensive view of the general behavior of
the instability, when one allows for non-parallel propagation
(i.e. $f \neq 0$, see also Gerwin \cite{ger}, Choudhury \&
Lovelace \cite{roy}), is to represent the results on a contour
plot such as the one in Fig. \ref{cont001}, where the contour
levels show the growth rate $\mathrm{Im}(\phi)$ as a function of
$f$ and $\Mr$, for fixed Alfv\`en Mach number $\zeta=1.6$ (top
panel). From the plot, one derives that i) two separate
instability regions are present in the plane $(\Mr,f)$, corresponding to slow
and fast magnetosonic unstable modes, and ii)
they do not extend to the horizontal axis, i.e. at $f=0$. By
expanding the solution of the dispersion relation in terms of
$1/f$, one can show from Eqs. (\ref{first}, \ref{second}) that the
corresponding first order term of the growth rate can be expressed
by:
\begin{equation}
\label{gr_asy} \phi \sim \frac{1}{f}
\left[\frac{V_{\mathrm{A}}^2-\beta^2}{1-\beta^2V_{\mathrm{A}}^2}\right]^{\frac{1}{2}}.
\end{equation}

Eq. (\ref{gr_asy}) demonstrates that, for large values of $f$, the instability
appears when $V_{\mathrm{A}} < \beta$.
In Fig. \ref{cont001} (bottom panel) we
show the contour plot of constant $\mathrm{Im}(\phi)$ surfaces as
a function of $f$ and $\beta$ for $V_{\mathrm{A}} = 10^{-3}$ and
$C_{\mathrm{s}} = 6.3 \times 10^{-4}$ (this value corresponds to $\Mr = 1.6$ (top
panel), see Eq. \ref{Mach}), and the figure shows that the instability appears
for $\beta>10^{-3}$, i.e. when $\beta$ exceeds the value of Alfv{\'e}n speed
parameter.
We note as well that, while the result in Eq. (\ref{gr_asy}) has a general validity,
in the non relativistic limit  $\beta \ll 1$,
 the condition for instability
derived above for large $f$'s reads also $\zeta < \Mr$, as evident for
Fig. \ref{cont001} (top panel).

For smaller $f$'s, i.e.
for small inclinations of the propagation direction with respect
to the velocity interface, Fig. \ref{cont001} shows that
for $f  \lapp  0.2$ the system becomes stable for the set of parameters considered.

\begin{figure}
\resizebox{\hsize}{!}{\includegraphics{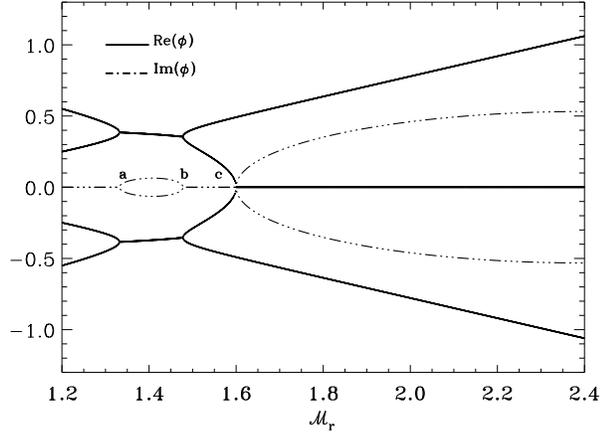}}
 \caption{Imaginary (dotted-dashed lines) and real
          (solid lines) parts of the roots satisfying
          the dispersion relation as functions of the relativistic
          Mach number. Only roots associated with unstable modes
          are plotted. For the sake of clarity, unphysical roots
          have been plotted as well.
          The other parameter set: $\beta=10^{-3}$, $\zeta=1.6$, $f=1$ and $\Gamma=5/3$.
          Instability arises when $a < \Mr < b$ and $\Mr > c$. }
\label{reimag}
\end{figure}

This general behavior is better understood by looking at the real and imaginary
parts of the roots giving rise to unstable modes, shown
in Fig. \ref{reimag}.
This plot corresponds to a horizontal cut across the contour
lines of Fig. \ref{cont001} for fixed $f = 1$.
Indeed, for $\Mr < a$, all roots are real yielding
a stable configuration.
For $a < \Mr < b$, however, the roots become complex conjugate
with nonzero imaginary parts, making this range unstable.
Between $b < \Mr < c$ we have again a stable configuration,
whereas for $\Mr > c$ the roots become purely imaginary and a
second unstable region can be recognized.
It can be verified by direct substitution
into Eq. (\ref{ssonic}) and Eq. (\ref{fsonic}) that the two family
of solutions separately correspond to the onset of slow (for $a <
\Mr < b$) and fast ($\Mr > c$) magnetosonic modes, respectively.

\begin{figure}
\resizebox{\hsize}{!}{\includegraphics{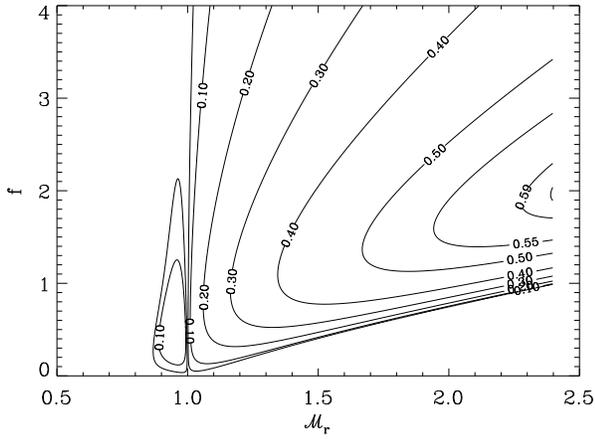}}
 \caption{Same as Fig. \ref{cont001}, but for $\zeta = 1$.}
 \label{cont001_1.2}
\end{figure}

For decreasing magnetization (i.e. lower values of $\zeta$), the
two instability regions approach each other, gradually shifting
towards lower values of $\Mr$. At the same time, lower
values of $f$ become prone to instability, see Fig. \ref{cont001_1.2}.
The limiting case $f=0$ is reached for $\zeta<1$,
when the previously identified instability ranges merge into a single
unstable region.

\begin{figure}
\resizebox{\hsize}{!}{\includegraphics{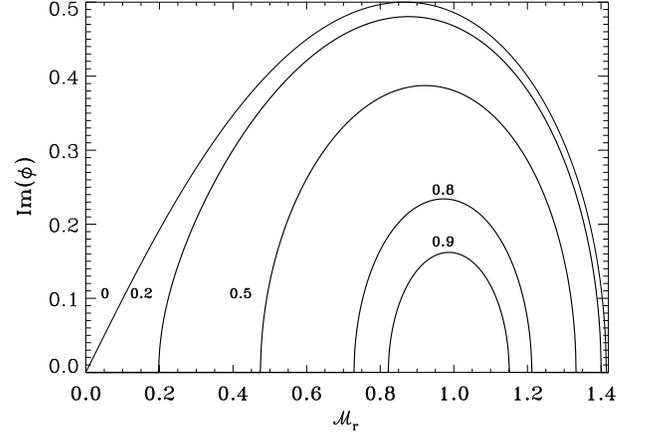}}
 \caption{Dependence of $\mathrm{Im}(\phi)$ on the relativistic Mach number
          for fixed $\beta = 10^{-3}$ and $f = 0$ ($\Gamma = 5/3$).
          The different labels above each curve give the values of the Alfv{\'e}nic
          Mach number $\zeta = [0,0.2,0.5,0.8,0.9]$.}
 \label{beta001}
\end{figure}

We will now examine the case of parallel propagation ($f=0$) for
different values of $\zeta<1$. These results may be directly compared with
those obtained by Ferrari et al. \cite{fer}.
In Fig. \ref{beta001} we plot the growth rate as a function of
$\Mr$ for several values of the Alfv{\'e}n Mach number
in the range $0\le \zeta \le 0.9$. For vanishing magnetic field
($\zeta = 0$) the instability disappears when $\Mr >
\sqrt{2}$, consistently with the results shown by Bodo et al. (\cite{bodo04}).
As the magnetic field increases, higher values of $\zeta$ induce a stabilizing
effect by raising the pressure and forcing the flow to be channeled along
the field lines. This has two noticeable effects: one is to decrease the
maximum value attained by the growth rate and the other is narrowing down
the instability Mach range $\Delta\Mr\equiv \Mr^{\max}-\Mr^{\min}$, where
$\Mr^{\max}$ ($\Mr^{\min}$) is the largest (smallest) value of the Mach number
above (below) which the configuration is stable. Note that the values of
the sound speed at the growth rate maxima are not relativistic.

\begin{figure}
\resizebox{\hsize}{!}{\includegraphics{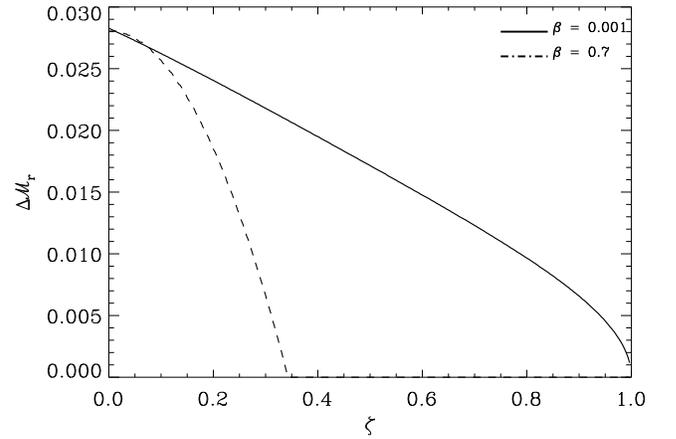}}
\caption{Instability Mach range $\Delta\Mr$ as a function of the
magnetic field in the case of parallel propagation for $\beta =
10^{-3}$ (solid line, $\Gamma = 5/3$) and $\beta = 0.7$ (dashed
line, $\Gamma = 4/3$). For the sake of comparison, the latter has
been reduced by a factor of $50$. Clearly, the instability is
suppressed when the Alfv{\'e}nic Mach number equals to $1$ (for
$\beta = 10^{-3}$) and $0.35$ (for $\beta = 0.7$).} \label{mczt}
\end{figure}

Fig. \ref{mczt} shows the dependence of the critical range of
the relativistic Mach number on the magnetic field (solid line).
As it is clear from the plot, when $\zeta\rightarrow 1$, the
relativistic Mach number range tends to zero and the instability
disappears. For $\zeta > 1$ the interface is always
stable. Our results are thus in full agreement with those of
Ferrari et al. \cite{fer} for the same parameter range ($\beta =
10^{-3}$, $f = 0$).

\subsection{Relativistic flows} \label{sec:rel}
%
%
%
%
%
%

In the previous subsection all the relevant quantities were
describing essentially flows in non-relativistic conditions.
We now consider situations in which the fluid exhibits
relativistic behavior, at least in some parameter range.


\begin{figure}
\resizebox{\hsize}{!}{\includegraphics{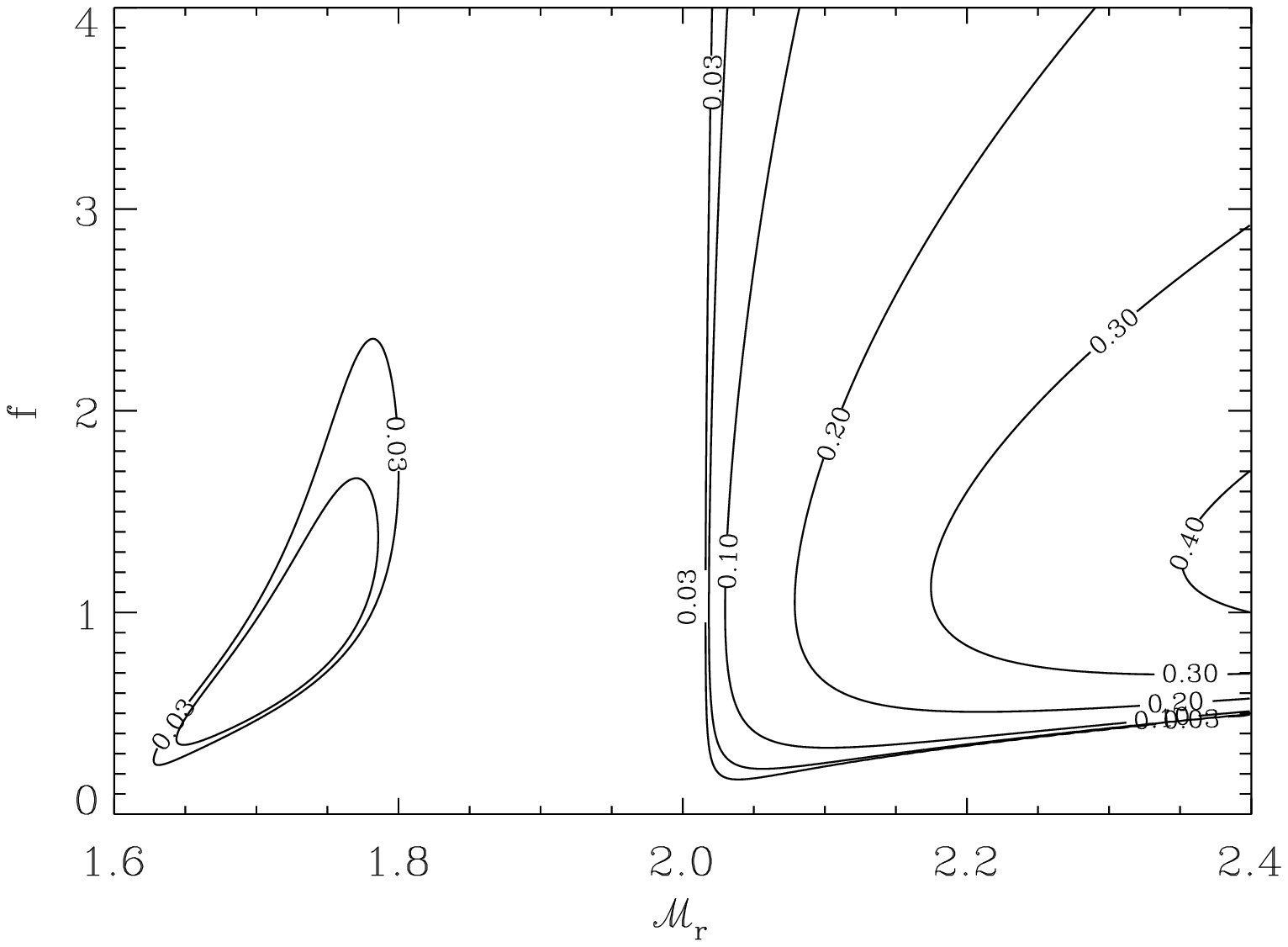}}
\resizebox{\hsize}{!}{\includegraphics{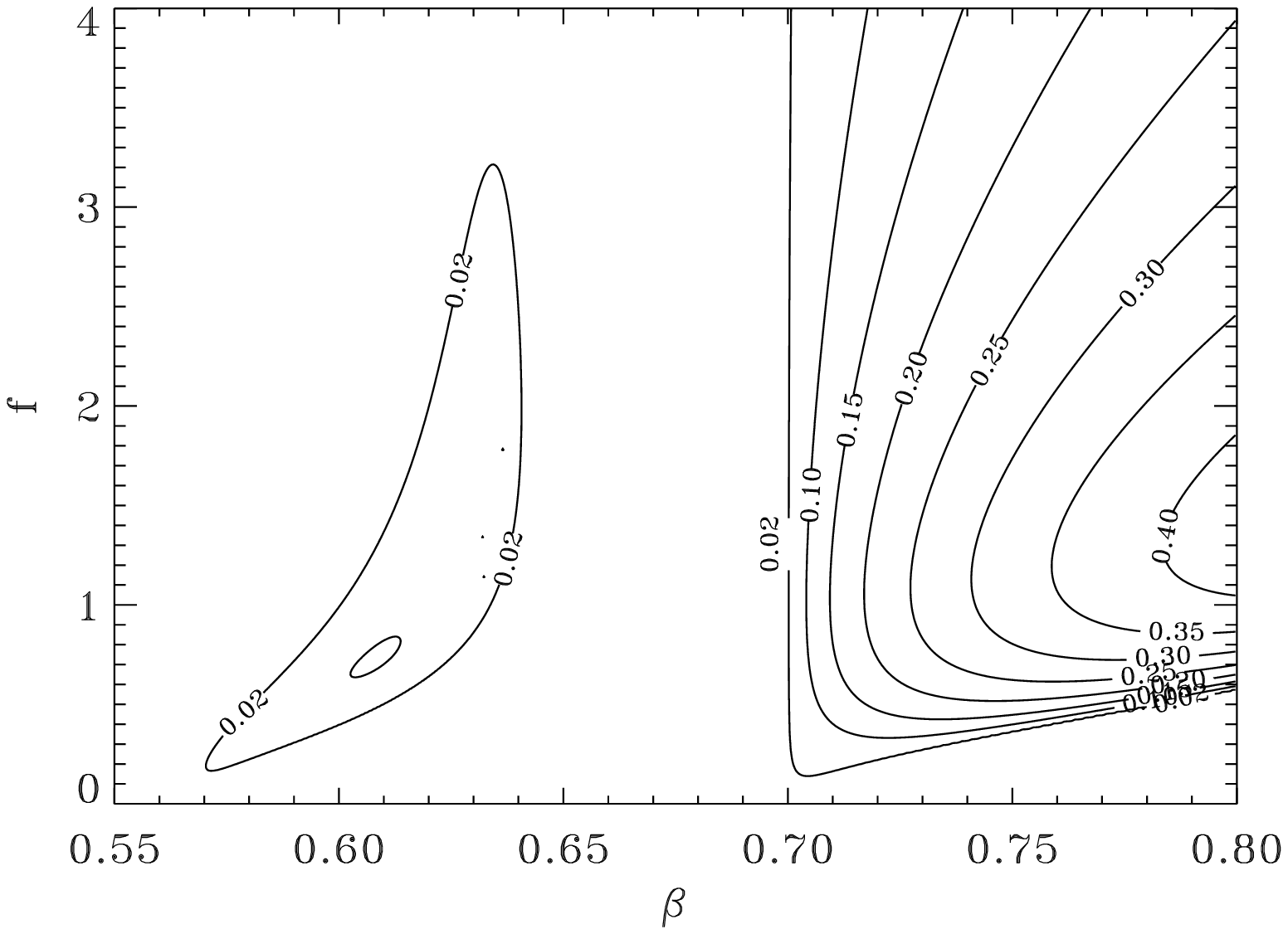}}
 \caption{Top panel: contour plots of constant $\mathrm{Im}(\phi)$ lines
          in the $\Mr-f$ plane for $\beta = 0.7$, $\zeta
          =1.6$. The two instabilities branches are clearly visible. Bottom panel: contour plots of constant $\mathrm{Im}(\phi)$ lines
          in the $\beta-f$ plane for $V_{\mathrm{A}} = 0.7$, $C_{\mathrm{s}} =
          0.43$. Here $\Gamma = 4/3$ was used.} \label{cont07}
\end{figure}

\begin{figure}
\resizebox{\hsize}{!}{\includegraphics{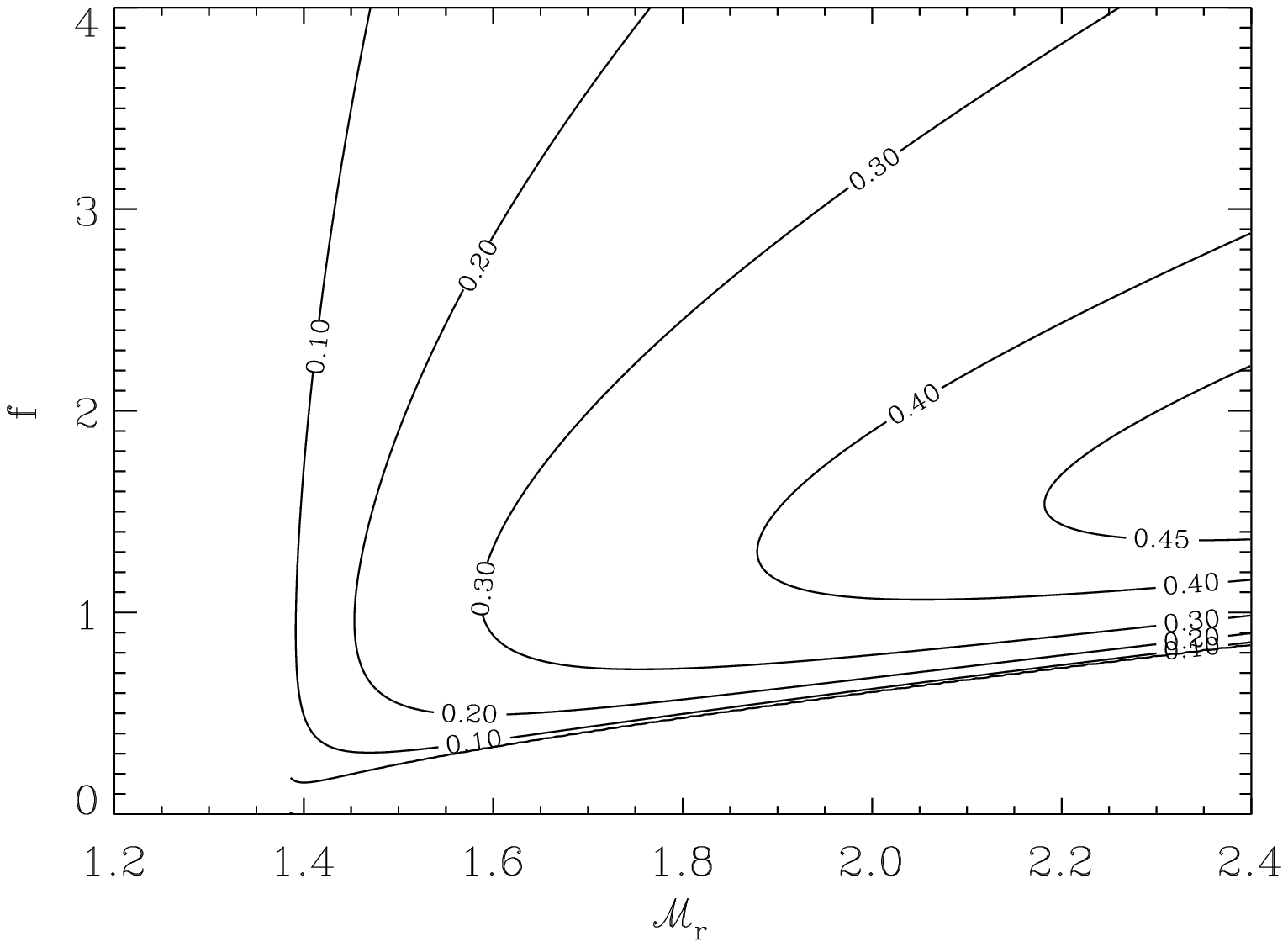}}
\resizebox{\hsize}{!}{\includegraphics{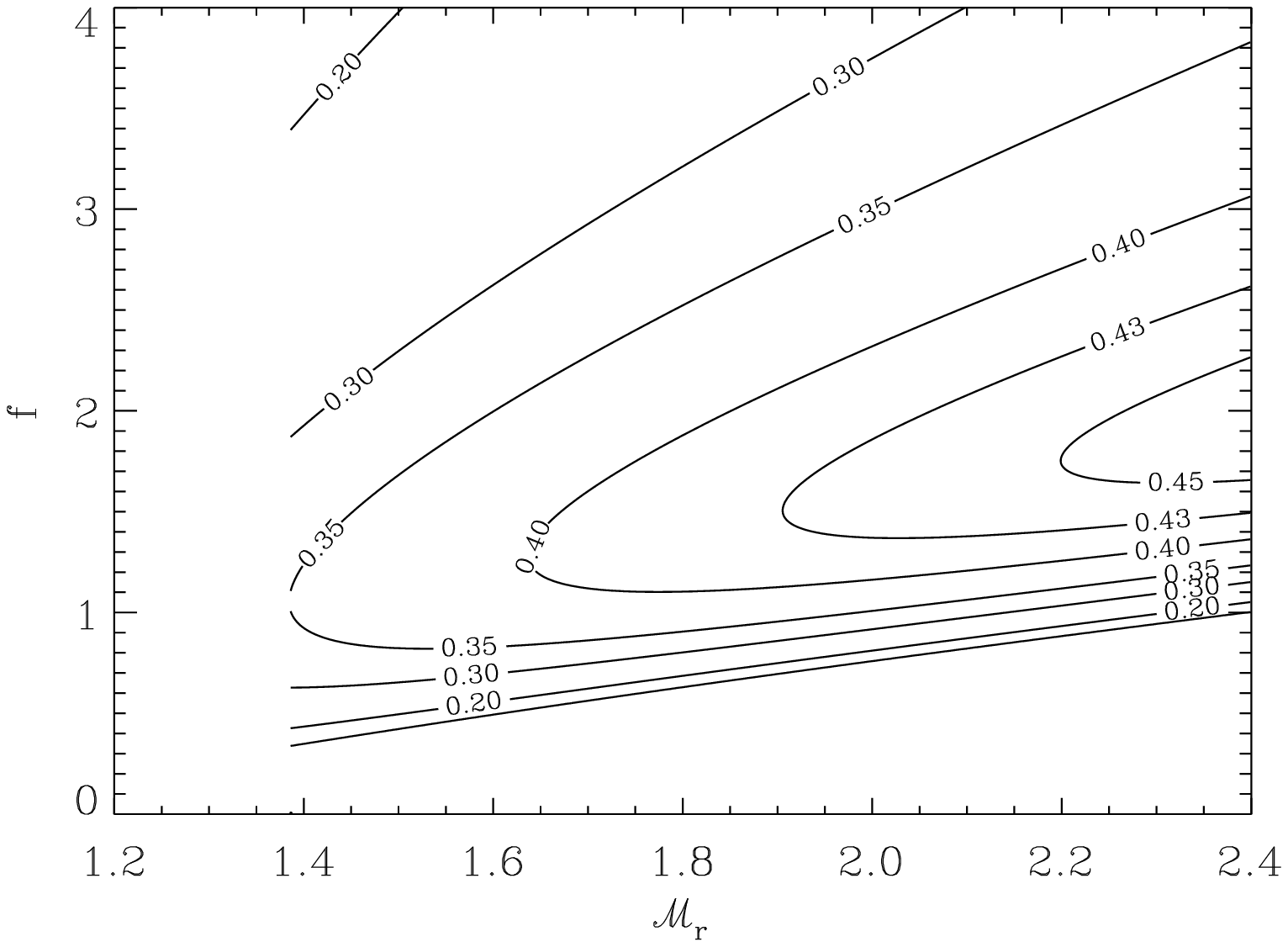}}
 \caption{Contour plots of constant $\mathrm{Im}(\phi)$ lines
          in the $\Mr-f$ plane for $\beta = 0.7$, $\zeta =1.2$ (top
          panel) and $\zeta = 1$ (bottom panel).
          Here $\Gamma = 4/3$ was used.
          }
\label{cont07_1}
\end{figure}

The top panel in Fig. \ref{cont07} shows the contour levels of the
growth rate for fixed flow velocity $\beta = 0.7$ and Alfv{\'e}nic
Mach number $\zeta=1.6$.   We have
again a slow magnetosonic modes instability region
for $1.63< \Mr < 1.8$, and a fast magnetosonic modes instability one for
$\Mr>2.02$. For large $f$'s the latter one only survives
and the discussion for the non relativistic case
is still valid, i.e. the instability condition is $V_{\mathrm{A}}<\beta$
(see Fig. \ref{cont07}
bottom panel). In fact, a direct comparison between Figs. \ref{cont07}
and \ref{cont001} reveals a similar qualitative behavior between
the two instability regions corresponding to slow and fast
magnetosonic modes. Indeed, in the relativistic case, higher
values of $\Mr$ are necessary to promote instability and the gap
separating the two regions becomes larger.
However, for lower values of $\zeta$, only one instability region
survives and there are no contour lines for $\Mr < 1.4$, see the
top panel in Fig. \ref{cont07_1}. This is a consequence of the
fact that higher values of the relativistic Mach number correspond
to lower sound speeds and, since the latter cannot exceed the
upper limiting value of $1/\sqrt{3}$, $\Mr$ has a lower physical
cutoff at $\sqrt{2}\beta\gamma\simeq 1.386$. Thus, no merging can
take place as $\zeta$ decreases, since the leftmost region
disappears below this threshold. This effect is clearly visible in
Fig. \ref{cont07_1}.
\begin{figure}
\resizebox{\hsize}{!}{\includegraphics{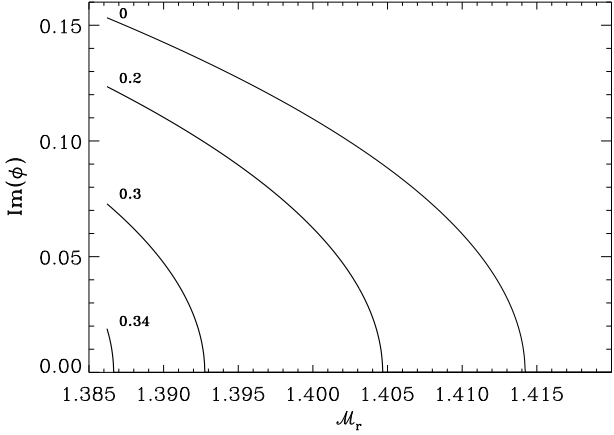}}
 \caption{Effects of magnetic field on the growth rate.
          Labels on each curve indicate values of
          $\zeta$. The set of parameters is: $\beta = 0.7$, $\zeta =
          [0, 0.2, 0.3, 0.34]$, $f = 0$ and $\Gamma = 4/3$. For $\zeta=0.35$
          the flow is completely stabilized.}
\label{beta7}
\end{figure}
In the limit of parallel propagation ($f = 0$), the growth rate
decreases with stronger magnetization, see Fig. \ref{beta7}.
This is also manifested by a reduced range of the Mach number
values for which instability exists.
This behavior has already been discussed for
non-relativistic flows in \S\ref{sec:nonrel}, where the flow was
shown to become stable for values of the Alfv{\'e}n speed close to
the speed of sound. For the present case ($\beta=0.7$), however,
stability is approached when the Alfv{\'e}n speed
$V_{\mathrm{A}}\simeq 0.35C_{\mathrm{s}}$, see Fig. \ref{beta7}.
This tendency can also be recognized by the profile of $\Delta\Mr$
(the critical Mach number instability range) as function of
$\zeta$ shown in Fig. \ref{mczt} (dashed line). As anticipated,
the instability is quenched for $\zeta\simeq 0.35$, an effect which
can also be justified by the increased kinematic inertia.

These results confirm the general trend already discussed in \cite{fer}; however
differences in the growth rates are found when the Alfv{\`e}n velocity approaches
the velocity of light and displacements currents become relevant.

\begin{figure}
\resizebox{\hsize}{!}{\includegraphics{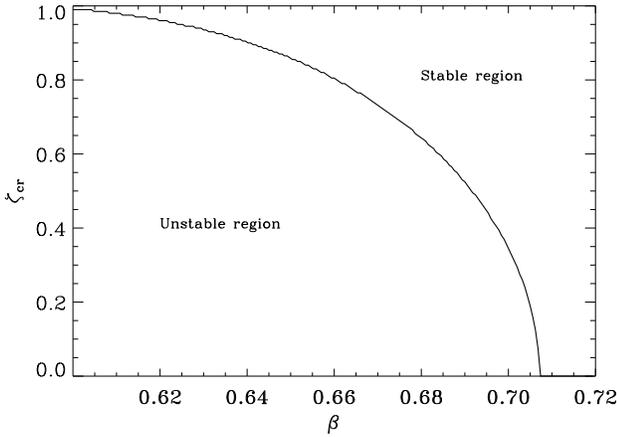}}
\caption{Effect of the kinematic factor on the
         instability when $f=0$ ($\Gamma = 4/3$).
         As it is clear, for $\beta\simeq 0.71$ the
         instability is suppressed by relativistic effects
         independently on $\Mr$.}
\label{btzt}
\end{figure}
In Fig. \ref{btzt}, indeed, we show the critical value of $\zeta$
(above which the instability is suppressed) as a function of the
flow velocity $\beta$. Thus $\zeta_{\rm cr}$ monotonically decreases
from $\sim 1$ (at $\beta = 0.6$) to $0$ for $\beta \simeq 0.71$.
Higher value of $\beta$ leads to a stable interface, even without
magnetic fields. This is a major difference from the classical MHD
case. The latter result was already introduced by Bodo et al. (\cite{bodo04}),
where it was shown that positive growth rates are subject to the
condition
\begin{equation}
\label{maxbeta}
\beta<C_{\mathrm{s}}\left[\frac{2}{1+C_{\mathrm{s}}^2}\right]^{\frac{1}{2}}\,.
\end{equation}
Since the right hand side is a monotonically increasing function
of $C_{\mathrm{s}}$, we conclude that for $\beta\geq 0.7071$
($C_{\mathrm{s}}=1/\sqrt{3}$) the instability is suppressed by
kinematic effects only, regardless of the value of the magnetic
field.

\section{Summary} \label{sec:summary}
%
%
%
%
%
%

We carried out a linear stability analysis of the relativistic
magnetized Kelvin-Helmholtz instability problem in the vortex
sheet approximation. Solutions to the dispersion relation have
been sought in terms of four parameters giving the strength of the
magnetic field, the relativistic Mach number, the flow velocity
and the spatial orientation of the wave vector. The main results
of this analysis can be summarized as follows:
\begin{enumerate}
\item We have examined the cases of non-relativistic and relativistic flows
      separately. We have found that, for fixed values of the Alfv{\`e}n velocity,
      two separate regions of instability appear in the plane $({\cal M}_{\rm r},f$).
      These two instability regions are associated with the destabilization of slow and
      fast magnetosonic modes.  For large $f$'s, only the fast magnetosonic modes are unstable
in the region $V_{\mathrm{A}} < \beta$.
\item For high values of the Alfv{\`e}n velocity, modes propagating parallel to the
      flow velocity are stable. As this velocity decreases, the two instability
      regions corresponding to slow and fast magnetosonic modes
      tend to merge and the parallel modes eventually destabilize.
      This general behavior
      is common for both non-relativistic and relativistic flows.
\item We have found that slow modes non-parallel to the flow direction
      are gradually excluded from the instability plane
      as they become unphysical. This effect takes place
      for increasing relativistic velocities or low magnetic
      fields or a combination of both.
\item When considering the case of parallel propagation ($f=0$),
      we have found that, similar to the non-magnetic counterpart, the flow
      becomes linearly stable when the relativistic Mach number exceeds
      a critical value. In the limit of vanishing Lorentz force, this threshold reaches
      the maximum value of $\sqrt{2}$, in a frame where the fluids have
      equal and opposite velocities.

\item For increasing magnetic field strength, the maximum unstable
      growth rate decreases and the instability exists for a narrow
      range of Mach number values.

\item The main difference from the non relativistic MHD is that,
      at higher flow velocities (for $f=0$),
      kinematic effects stabilize the flow even for smaller values of
      the Alfv{\'e}n velocity. Furthermore, the computed growth rates
      attain lower values than their classical counterparts. When the
      flow velocity becomes higher than $1/\sqrt{2}$, the flow does not
      need any more the magnetic field for stabilization.
\end{enumerate}
      Relativistic flows stabilize the KH modes due to the concurrence
      of two effects: first, at high relative velocities, the mode coupling between the
      two half-spaces loosens because of causality effects, this effect is
      present in the non-relativistic case as well but is even more
      relevant at relativistic velocities; second, in the relativistic regime the inertia of the fluid particles
      is dynamically augmented, hampering the response of the fluid to the instability.

Destabilization of relativistic Kelvin-Helmholtz modes has been applied
in the astrophysical context of jets of extended extragalactic
radiosources and relativistic galactic sources to explain the generation
of observed morphologies and the driving mechanisms for their radiation
emission. While originally these modes were thought to yield disruption
of the collimated propagation, it was later realized that the head of
jets creates a bow shock and a cocoon that shield the flow from steep
gradient boundary layers while leading to generation of a turbulent state.
Turbulence is functional in producing
mixing of jet and ambient matter with entrainment, angular momentum transport
and slowing down of the flow (Rossi et al. 2008). The behavior of KHI as a function of the
physical parameters is likely to the at the base of the different classes of radiosources.
In addition long wavelength modes  may develop and generate
global structures as the wiggles and knots that are observed in
astrophysical jets.
The present fully relativistic analysis confirms that in order to activate
KH modes small $V_A / C_S$ ratios are needed.
This condition can be realized in particular by very hot plasmas,
even for relatively large magnetic fields, as it appears to be
the case in high-energy sources.

KHI appears also to be relevant in the case of PWNe; in fact in these objects a relativistically
hot magnetized plasma is produced by the pulsar ultra-relativistic wind and
interacts with the surrounding supernova ejecta. These
PWNe are internally structured and presents high
velocity flow channels that can become KH unstable.
As already mentioned, Bucciantini \& Del Zanna
\cite{buc} have examined numerically the nonlinear development of KHI in
these channels, assuming planar geometry, for interpreting the signature of the instability
on the synchrotron emission.
The present linear stability analysis is complementary to their
study as it allows a better comprehension of the physical phenomena involved
in the destabilization process. Again the presence of a hot plasma appears necessary to start the
channel destabilization.

\begin{acknowledgements}
   The authors are grateful to Dr. Edoardo Trussoni for valuable
   discussions. The research was partially supported by the Georgian National Science
   Foundation grant GNSF/ST06/4-096.\end{acknowledgements}


\begin{thebibliography}{}

\bibitem[1989]{anile} Anile A. M., 1989, {\it Relativistic fluids and Magneto-Fluids}, Cambridge Univ. Press, Cambridge

\bibitem[1976]{bln}  Blandford R. \&  Pringle J., 1976, Mon. Not. R. Astron. Soc., {\bf 176}, 443

\bibitem[2003]{bodo03} Bodo, G., Rossi, P., Mignone, A., Massaglia, S., \& Ferrari, A.\ 2003,
                       New Astronomy Review, 47, 557

\bibitem[2004]{bodo04} Bodo G., Mignone A. \& R. Rosner, 2004, Phys.Rev.E {\bf 70},036304

\bibitem[2006]{buc} Bucciantini  N. \& Del Zanna L., 2006, A\&A, {\bf
    454}, 393

\bibitem[Chandrasekhar 1961]{chandra} Chandrasekhar, S., 1961, {\it Hydrodynamic and Hydromagnetic Stability}, Oxford: Clarendon

\bibitem[1987]{hardee87} Hardee, P.~E. 1987, ApJ , 313, 607

\bibitem[2007]{hardee07} Hardee, P.~E. 2007, ApJ , 664, 26

\bibitem[1980]{fer} Ferrari A., Trussoni E., \& Zaninetti L., 1980, Mon. Not. R. Astron. Soc., {\bf 193}, 469

\bibitem[1981]{fer81}  Ferrari A., Trussoni E., \& Zaninetti L., 1981, Mon. Not. R. Astron. Soc., {\bf 196}, 1051

\bibitem[1982]{fer82}  Ferrari A., Massaglia, S. \& Trussoni E., 1982, Mon. Not. R. Astron. Soc., {\bf 198}, 1065

\bibitem[Gerwin 1968]{ger}  Gerwin R.A., 1968, Rev. Mod. Phys. {\bf 40}, 652


\bibitem[1999]{kom} Komissarov S.S., 1999, Mon. Not. R. Astron. Soc., {\bf 303}, 343

\bibitem[1980]{kon80} K\"onigl, A., 1980, Phys. Fluids, {\bf
  23}, 1083.

\bibitem[1871]{kel} Lord Kelvin, 1871, Philos. Mag. {\bf 42}, 362

\bibitem[1967]{lichne} Lichnerowicz, A.\ 1967,
    {\it Relativistic Hydrodynamics and Magnetohydrodynamics}, New York: Benjamin,

\bibitem[1999]{mir} Mirabel, I.F., Rodriguez, L.F., 1999, ARA\&A, {\bf 37}, 409

\bibitem[Pu \& Kivelson 1983]{PK83} Pu, Z.Y., \& Kivelson, M.G., 1983,
  J.Geophys.Res., {\bf 88}, 841

\bibitem[1986]{roy} Roy S. Choudhury, \& Lovelace R.V., 1986, ApJ, {\bf 302}, 188


\bibitem[2004]{rossi04} Rossi, P., Bodo, G., Massaglia, S., Ferrari, A., \& Mignone, A.
                        2004, ApSS, 293, 149

\bibitem[Rossi 2008]{rossi08} Rossi, P., Mignone, A., Bodo, G., Massaglia, S., \& Ferrari, A.\
                        2008, A\&A, in press

\bibitem[Sen 1963]{sen63} Sen, A.K., 1963, Phys. Fluids, {\bf 6}, 1154

\bibitem[Sen 1964]{sen64} Sen, A.K., 1963, Phys. Fluids, {\bf 7}, 1293

\bibitem[1976]{tur}  Turland B.D. \&  Scheuer P.A.G., 1976, Mon. Not. R. Astron. Soc., {\bf 176}, 421

\bibitem[1868]{hel} Von Helmholtz H., Monats K., 1868, Preuss. Akad. Wiss. Berlin {\bf 23}, 215







\end{thebibliography}
\end{document}